\documentclass[%
reprint,
showpacs,
nofootinbib,
amsmath,amssymb,
aps,
pra,
]{revtex4-1}

\usepackage{graphicx}
\usepackage{dcolumn}
\usepackage{bm}
\usepackage{physics}
\usepackage{suffix}
\usepackage{xstring}
\usepackage[usenames, dvipsnames]{color}
\usepackage{gitinfo}
\usepackage{siunitx}
\DeclareSIUnit\Gauss{G}
\DeclareSIUnit\bohr{bohr}
\usepackage[backgroundcolor=blue!20!white]{todonotes}
\usepackage[normalem]{ulem}

\newcommand*{\gF}[1]{\ensuremath{g_f^{#1}}}

\newcommand*{\muB}{\ensuremath{\mu_\mathrm{B}}}
\newcommand*{\dressedState}[0]{\ensuremath{\widetilde{m}}}
\newcommand*{\manifold}{\ensuremath{m_\text{tot}}}
\newcommand*{\hyperfineState}[0]{\ensuremath{f}}
\newcommand*{\hyperfinem}[1]{\ensuremath{m_{f#1}}}

\newcommand*{\Brf}[0]{\ensuremath{B_\mathrm{\rf{}}}}
\WithSuffix\newcommand\Brf*{\ensuremath{\vec{B}_\mathrm{\rf{}}}}
\newcommand*{\Btop}[0]{\ensuremath{B_\mathrm{TA}}}
\WithSuffix\newcommand\Btop*{\ensuremath{\vec{B}_\mathrm{TA}}}
\newcommand*{\Bzero}[0]{\ensuremath{B_\mathrm{0}}}
\WithSuffix\newcommand\Bzero*{\ensuremath{\vec{B}_\mathrm{0}}}
\newcommand*{\drf}[0]{\ensuremath{\omega}}
\newcommand*{\Bquad}[0]{\ensuremath{B'}}
\newcommand*{\Beff}[0]{\ensuremath{\vec{V}}}
\WithSuffix\newcommand\Beff*{\ensuremath{B_\text{eff}}}

\newcommand*{\tiltAngle}[1]{\ensuremath{\theta_{\mathrm{#1}}}}
\newcommand*{\scTiltAngle}[1]{\ensuremath{\Theta_{\mathrm{#1}}}}

\newcommand*{\AN}[1]{\ensuremath{\mathcal{N}_{#1}}}
\newcommand*{\kTwo}[1]{\ensuremath{k_2^{\mathrm{#1}}}} 
\newcommand*{\kThree}[1]{\ensuremath{k_3^{\mathrm{#1}}}} 
\newcommand*{\density}[1]{\ensuremath{n_{#1}}}
\newcommand*{\Temperature}[1]{\ensuremath{T_{\mathrm{#1}}}}

\newcommand*{\rf}[0]{RF}
\newcommand*{\Rb}[1]{\ensuremath{\mathrm{^{#1}Rb}}}
\newcommand*{\Potassium}[1]{\ensuremath{\mathrm{^{#1}K}}}
\newcommand*{\Sodium}[1]{\ensuremath{\mathrm{^{#1}Na}}}
\newcommand*{\Caesium}[1]{\ensuremath{\mathrm{^{#1}Cs}}}
\newcommand*{\Lithium}[1]{\ensuremath{\mathrm{^{#1}Li}}}

\newcommand*{\etal}[0]{\emph{et al.}}
\newcommand*{\refform}[1]{%
	\IfBeginWith{#1}{eq:}{Eq.~\eqref{#1}}{}%
	\IfBeginWith{#1}{fig:}{Fig.~\ref{#1}}{}%
	\IfBeginWith{#1}{tab:}{Table~\ref{#1}}{}%
	\IfBeginWith{#1}{appendix:}{Appendix~\ref{#1}}{}%
	\IfBeginWith{#1}{sec:}{Section~\ref{#1}}{}%
}

\newcommand*{\coordaxis}[1]{\ensuremath{\vec{e}_{#1}}}
\newcommand*{\Hamiltonian}[1]{\ensuremath{\mathcal{H}_\text{#1}}}
	
\begin{document}
	
	\preprint{APS/123-QED}
	
	\title{Inelastic collisions in radiofrequency-dressed mixtures of ultracold atoms}
	
	\author{Elliot Bentine}%
		\email{elliot.bentine@physics.ox.ac.uk}
		\affiliation{%
		Department of Physics, Oxford University, Parks Road, OX1 3PU
		}%
	\author{Adam J. Barker}%
		\affiliation{%
		Department of Physics, Oxford University, Parks Road, OX1 3PU
		}%
	\author{Kathrin Luksch}%
		\affiliation{%
		Department of Physics, Oxford University, Parks Road, OX1 3PU
		}%
	\author{Shinichi Sunami}%
		\affiliation{%
		Department of Physics, Oxford University, Parks Road, OX1 3PU
	}%
	\author{Tiffany L. Harte}%
		\altaffiliation{Department of Physics, Cambridge University, CB3 0HE, UK}
		\affiliation{%
		Department of Physics, Oxford University, Parks Road, OX1 3PU
		}%
	\author{Ben Yuen}%
		\affiliation{%
		Department of Physics, Oxford University, Parks Road, OX1 3PU
	}%
	\author{Christopher J. Foot}%
	\affiliation{%
		Department of Physics, Oxford University, Parks Road, OX1 3PU
	}%
	\author{Daniel J. Owens}%
	\affiliation{%
		Joint Quantum Centre (JQC) Durham-Newcastle, Department of Chemistry,
		Durham University, South Road, Durham DH1 3LE
	}%
	\author{Jeremy M. Hutson}%
        \email{J.M.Hutson@durham.ac.uk}
		\affiliation{%
		Joint Quantum Centre (JQC) Durham-Newcastle, Department of Chemistry,
Durham University, South Road, Durham DH1 3LE
	}%
	
	\date{\today}
	
	\begin{abstract}
		Radiofrequency (\rf{})-dressed potentials are a promising technique for manipulating atomic mixtures, but so far little work has been undertaken to understand the collisions of atoms held within these traps.
		In this work, we dress a mixture of \Rb{85} and \Rb{87} with \rf{} radiation, characterize the inelastic loss that occurs, and demonstrate species-selective manipulations.
		Our measurements show the loss is caused by two-body \Rb{87}+\Rb{85} collisions, and we show the inelastic rate coefficient varies with detuning from the \rf{} resonance.
		We explain our observations using quantum scattering calculations, which give reasonable agreement with the measurements.
		The calculations consider magnetic fields both perpendicular to the plane of RF polarization and tilted with respect to it.
		Our findings have important consequences for future experiments that dress mixtures with \rf{} fields.
	\end{abstract}
	

	\maketitle

	Experiments that use mixtures of ultracold atoms are now established as versatile quantum simulators for a range of physical phenomena.
	For systems of many particles, recent studies have examined superfluidity~\cite{Ferrier-Barbut2014}, non-equilibrium dynamics in many-body quantum systems~\cite{Cetina2016}, and the interactions mediated by a bath~\cite{DeSalvo2019}. 
	At the single-particle level, experiments have observed diffusion~\cite{Hohmann2017}, chemical reactions~\cite{Liu2018} and ultralow-energy collisions~\cite{Schmidt2019}.
	Mixtures of ultracold atoms are also used as a starting point for the production of ultracold molecules~\cite{Papp2006,DeMarco2019}.
	These experiments have been made possible by techniques to manipulate ultracold mixtures and their constituents, and new investigations will become possible as laboratory methods evolve.	

	A number of different techniques are used to trap and manipulate cold atoms.
	In this paper we consider \rf{}-dressed potentials, which confine cold atoms through a combination of static and radiofrequency magnetic fields~\cite{GarrawayPerrinReview,Perrin2017}.
	Notable advantages of this technique include smooth, defect-free traps and low heating rates~\cite{Merloti2013}.
	The potential can be shaped by controlling the \rf{}-dressing field~\cite{Hofferberth2006}, or by adding additional \rf{} components~\cite{Harte2017}.
	Furthermore, the potential may be combined with additional time-averaging fields to produce trap geometries such as rings or double wells~\cite{Lesanovsky2007,Gildemeister2010,Sherlock2011}.
	The confining forces depend only on an atom's magnetic structure, unlike optical methods which depend on electronic structure.
	Therefore, \rf{}-dressed potentials permit species-selective manipulations of mixtures that have similar confinement in dipole traps, such as mixtures of hyperfine states or isotopes~\cite{Extavour2006,Navez2016,Bentine2017}.
	
	In spite of the advantages of \rf{}-dressed potentials, there has so far been little consideration of the collisional stability of mixtures that are trapped using them.
	In this work, we investigate collisions in an \rf{}-dressed mixture of \Rb{85} and \Rb{87} atoms in their lower hyperfine states.
	We observe a rapid loss of \Rb{85} atoms from the trap due to two-body \Rb{87}+\Rb{85} inelastic collisions, which occur through a spin-exchange mechanism.
	We use a theoretical model to explain the inelastic collisions, and compare predictions from quantum scattering calculations to our measurements of the two-body rate coefficients.
	Our results suggest that spin-exchange collisions will occur for most combinations of alkali-metal atoms in \rf{}-dressed potentials.
	Furthermore, our results verify our understanding of ultracold collisions in the presence of strong, resonant dressing fields.	
	
	This paper is structured as follows.
	In \refform{sec:APReview} we explain the dressed-atom picture.
	In \refform{sec:ExperimentalSequence} we describe the experimental procedure used to produce cold clouds of \Rb{85} and \Rb{87}, and we demonstrate species-selective manipulations.
	In \refform{sec:Results} we present measurements of the inelastic loss, which we show is dominated by two-body \Rb{87}+\Rb{85} collisions,
	and we measure the two-body rate coefficient \kTwo{85,87} as a function of magnetic field and \rf{} field amplitude.
	In \refform{sec:QuantumScatteringCalculations} we discuss the quantum scattering calculations and compare the predicted rate coefficients to the experimental measurements.
	In \refform{sec:semiclassical} we compare our results to a semi-classical model.
	In \refform{sec:Conclusion} we conclude with a discussion of inelastic loss expected for other species.
	
	\section{The \rf{}-dressed atom}
	\label{sec:APReview}
	
	In this work, we consider rubidium atoms in their ground electronic state.
	We adopt the common collision convention that lower-case quantum numbers refer to individual atoms and upper-case quantum numbers refer to a colliding pair.
	An atom is described by its electron spin $s=1/2$ and nuclear spin $i$, which couple to form a resultant spin $f$.
	In a static magnetic field \Bzero*{} along the $z$ axis, the Hamiltonian for each atom is
	\begin{equation}
	\Hamiltonian{atom} = \zeta \hat i \cdot \hat s + \left( g_S \hat s_z + g_i \hat i_{z}\right) \muB{} \Bzero{},
	\label{eq:mon}
	\end{equation}
	where $\zeta$ is the hyperfine coupling constant, \muB{} is the Bohr magneton and $g_S$ and $g_i$ are electron and nuclear spin $g$-factors with the sign convention of Arimondo \etal{}~\cite{Arimondo1977}.
	At the low magnetic fields considered here, each atomic state splits into substates with a well-defined projection \hyperfinem{} of the total angular momentum \hyperfineState{} along \Bzero*{}. In this regime, \hyperfineState{} is nearly conserved but the individual projections $m_s$ and $m_i$ of $s$ and $i$ are not.
	At low fields, the field-dependent terms in \refform{eq:mon} may be approximated in the coupled basis $\ket{\hyperfineState, \hyperfinem{}}$ to 
	\begin{equation}
	\Hamiltonian{atom} = \gF{} \muB{} \Bzero{} \hyperfinem{},
	\end{equation}
	where substates \hyperfinem{} are separated in energy by the Zeeman splitting and \gF{} is the Land\'e $g$-factor.
	
	In addition to the static magnetic field, we consider a radiofrequency field with angular frequency \drf{} that is $\sigma_-$ polarized about the $z$ axis, with $\vec{B}_\textrm{\rf{}}(t)=\Brf{} [
	\coordaxis{x} \cos \drf{} t-\coordaxis{y} \sin \drf{} t]$.
	The Hamiltonian of the \rf{} field is
	\begin{equation}
	\Hamiltonian{rf} = \hbar \drf{} (N+N_0),
	\end{equation}
	where $\hat{N}=\hat{a}_-^\dagger\hat{a}_- -\langle \hat{a}_-^\dagger \hat{a}_-\rangle$ is the photon number with respect to the average photon number $N_0=\langle\hat{a}_-^\dagger\hat{a}_-\rangle$, and $\hat{a}_-^\dagger$ and $\hat{a}_-$ are photon creation and annihilation operators for $\sigma_-$ photons.
	For $\sigma_-$ polarization, the $N$ photons have angular momentum projection $M_N = -N$ onto the $z$ axis. 
	
	The Hamiltonian for the interaction of the field with an atom is
	\begin{equation}
	\Hamiltonian{int} =\frac{\mu_{\rm B} B_{\rm rf}}{2\sqrt{N_0}}
	\left[(g_S\hat s_+ + g_i\hat i_+)\hat a_-^\dagger + (g_S\hat s_- + g_i\hat
	i_-)\hat a_-\right], \label{eq:rf_int}
	\end{equation}
	where $\hat s_+$ and $\hat s_-$ are raising and lowering operators for the
	electron spin and $\hat i_+$ and $\hat i_-$ are the corresponding operators for
	the nuclear spin.
	
\begin{figure}
	\begin{center}
		\includegraphics[width=\columnwidth]{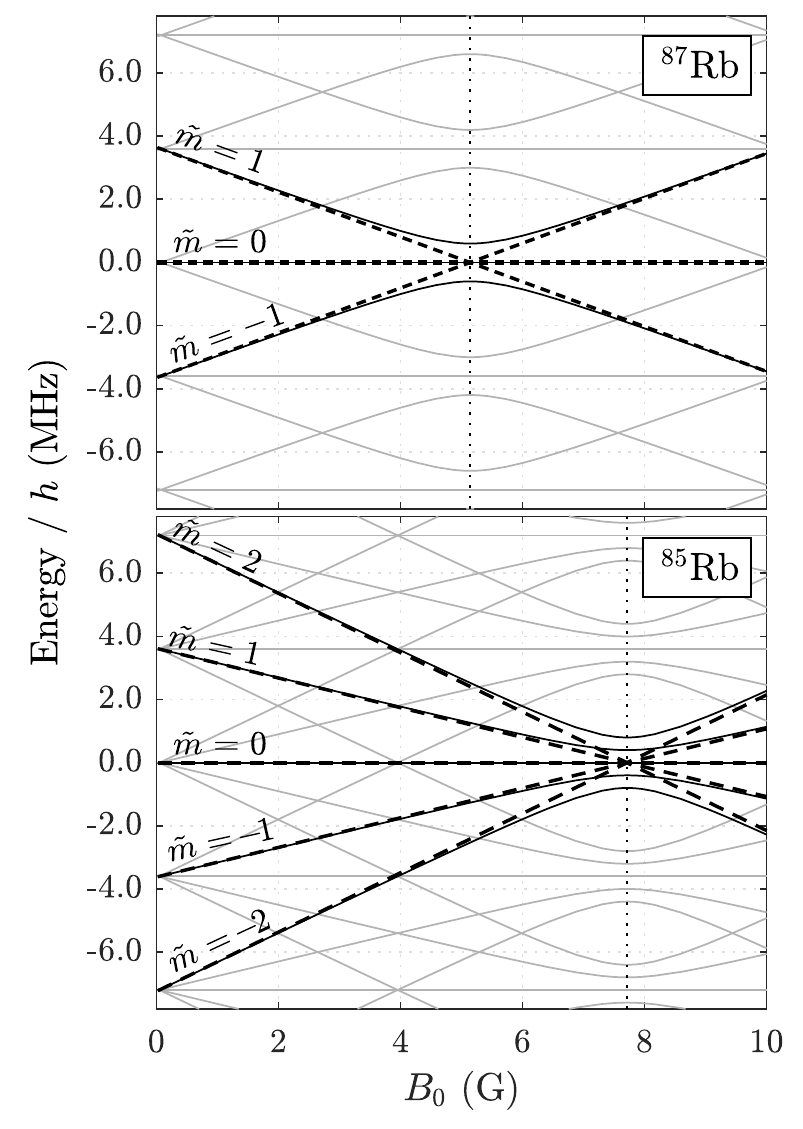}
	\end{center}
		\caption{The \rf{}-dressed eigenstates (solid lines) for \Rb{87} (top) and \Rb{85} (bottom) as a function of magnetic field \Bzero{}, for $\Brf{}=\SI{0.86}{\Gauss}$ and $\drf{}=\SI{3.6}{\mega\Hz}$.
		A single manifold with $\manifold{}=0$ is emphasized in bold.
		Dashed lines show the dressed states of this manifold in the limit of zero atom-photon interaction.
		\label{fig:dressedAtoms_Es}
	}
	\end{figure}

	The atom-photon interaction for $\sigma_-$ polarization conserves $\manifold{} = \hyperfinem{}+ M_N = \hyperfinem{} - N$.
	If the couplings are neglected, states with the same $\manifold{}$ cross as a function of magnetic field at the radiofrequency resonance $\hbar \omega = \gF{}\muB{}\Bzero{}$, as shown by the dashed lines in \refform{fig:dressedAtoms_Es}.
	For the \rf{} frequency of \SI{3.6}{\mega\Hz} used in this work, the states
	$(f,m_f,N) = (1,+1,1)$, $(1,0,0)$ and $(1,-1,-1)$ of \Rb{87} all cross near
	$B=\SI{5.12}{\Gauss}$, and the states $(f,m_f,N) = (2,+2,2), (2,+1,1)$, $(2,0,0)$,
	$(2,-1,-1)$ and $(2,-2,-2)$ of \Rb{85} all cross near $B=\SI{7.70}{\Gauss}$.
	These crossings become avoided crossings when the couplings of \refform{eq:rf_int} are included.
	The eigenstates within each manifold of constant $m_\text{tot}$ are labelled by the quantum number $\dressedState$, which takes values in the range $-\hyperfineState$ to $\hyperfineState$.
	The corresponding eigenenergies are
	\begin{equation}
	E(\dressedState{},\manifold{}) = \hbar \dressedState{} \sqrt{\Omega^2+\delta^2} - \manifold{} \hbar \omega,
	\end{equation}
	where $\delta = (\gF{} \muB{} \Bzero{} / \hbar - \drf{})$ is the angular frequency detuning from resonance and $\Omega$ is the Rabi frequency on resonance.
	In an inhomogeneous field, atoms in states for which $\dressedState > 0$ may be trapped in the resulting potential minimum~\cite{GarrawayPerrinReview,Perrin2017}.
	
	\section{Experimental Methods}			
	\label{sec:ExperimentalSequence}
	
	In this section, we describe the methods used to cool and trap mixtures of \Rb{85} and \Rb{87}.
	Our apparatus was described previously~\cite{Harte2017}, and has since been modified to allow the trapping of two species.
	
	An experimental sequence begins by collecting atoms of \Rb{85} and \Rb{87} into a dual-isotope magneto-optical trap (MOT).
	The cooling and repumping light for \Rb{87} is generated by two external-cavity diode lasers.
	Each laser is locked to one of the transitions, and injection-locks a laser diode that is current-modulated at a frequency of \SI{1.1}{\giga\Hz} (cooling) or \SI{2.5}{\giga\Hz} (repumper).
	The modulation generates sidebands at the frequencies required to laser cool and repump \Rb{85} atoms.
	Light from the two injection-locked diodes is combined and passed through a tapered amplifier, before illuminating a 3D pyramid MOT,
	which collects \num{4e9} atoms of \Rb{87} and \num{1e8} atoms of \Rb{85}.
	These atoms are optically pumped into their lower hyperfine levels, with $\hyperfineState{}=1$ and $2$ respectively, and the low-field-seeking states are loaded into a magnetic quadrupole trap.
	
	The trapped mixture of isotopes is transported to an ultra-high-vacuum region where it is evaporatively cooled using a weak \rf{} field, first within a quadrupole trap and then in a Time-Orbiting Potential (TOP) trap, to a temperature of $\sim\SI{0.5}{\micro\K}$.
	This process predominantly ejects \Rb{87} atoms from the trap and the \Rb{85} atoms are sympathetically cooled with minimal loss~\cite{Bloch2001}.
	The final atom numbers of each species are controlled by adjusting the power of the cooling light that is resonant with each isotope during the MOT loading stage, which determines the number of atoms initially collected.
	We observe no evidence of interspecies inelastic loss in these magnetic traps, imposing a bound of $\kTwo{85,87} \ll \SI{e-14}{\cm\cubed\per\s}$ on the two-body rate coefficient;
	this is expected because spin exchange is forbidden between the $(\hyperfineState{},\hyperfinem{})=(2,-2)$ and $(1,-1)$ states of \Rb{85} and \Rb{87} respectively.
	
	\subsection{Species-selective manipulations}
	\label{sec:SSTAAP}
	
	After evaporation, the atoms are loaded into a time-averaged adiabatic potential (TAAP)~\cite{Lesanovsky2007,Gildemeister2010}.
	This potential is formed by combining a spherical quadrupole field  $\vec{B}_\text{quad}$, a slow time-averaging field \Btop*{}, and an \rf{}-dressing field $\Brf*{}$ that is $\sigma_-$ polarized in a plane perpendicular to $z$:
	\begin{align}
		&\vec{B}_\text{quad}&=& B'(x \coordaxis{x} + y \coordaxis{y} - 2 z \coordaxis{z}), & \\
		&\Brf*{}&=& B_\text{rf} \left[\cos{(\omega t)} \coordaxis{x} - \sin({\omega t}) \coordaxis{y}\right], & \\	
		&\Btop*{}&=& \Btop{} \left[ \cos(\omega_\mathrm{TA} t) \coordaxis{x} - \sin(\omega_\mathrm{TA} t) \coordaxis{y} \right]. & 
	\end{align}
	The \rf{} field, with $\omega = \SI{3.6}{\mega\Hz}$, drives transitions between the Zeeman substates so that the atoms are in \rf{}-dressed eigenstates.
	The \rf{} field is resonant with the atomic Zeeman splitting at points on the surface of a spheroid, centered on the quadrupole node, with semi-axes of length $\hbar \omega / \gF{} \muB{} B' \times \lbrace 1, 1, 0.5 \rbrace$ along the $\lbrace\coordaxis{x},\coordaxis{y},\coordaxis{z}\rbrace$ axes.
	The time-averaging field sweeps this resonant surface in a circular orbit of radius $r_\mathrm{orbit} = \Btop{} / \Bquad{} $ around the $z$ axis.
	The frequency of the time-averaging field, $\omega_\mathrm{TA}=\SI{7}{\kilo\Hz}$, is slow compared to the Larmor, \rf{} and Rabi frequencies, so atoms adiabatically follow the \rf{}-dressed eigenstates as the potential is swept.
	
	For a single species, the TAAP operates in two modes, depending on the value of \Btop{}.
	When $\Btop{} > \hbar{} \drf / \gF{}\muB{}{}$, the resonant spheroid orbits far from the atoms, which are confined near the origin by the rotating field \Btop*{}, as in a TOP trap.
	When $\Btop{} < \hbar{}\drf{} / \gF{}\muB{}$, the resonant spheroid intersects the $z$ axis, forming a double-well potential with minima at positions~\cite{Lesanovsky2007}
	\begin{equation}
	\label{eq:TAAPminpos}
	x=0, y=0, z = \pm \frac{\hbar \omega}{2 \gF{} \muB{} \Bquad{}} \sqrt{1 - \left(\frac{\gF{}\muB{}\Btop{}}{\hbar \omega}\right)^2}.
	\end{equation}
	In this work we load atoms only into the lower well, and henceforth neglect the upper well.
	The vertical position of the lower well is controlled by changing \Btop{}, which determines the radius of orbit and thus the point of intersection of the resonant spheroid and the $z$ axis.

	\begin{figure}
	\begin{center}
		\centering{}
		\includegraphics[width=0.45\textwidth]{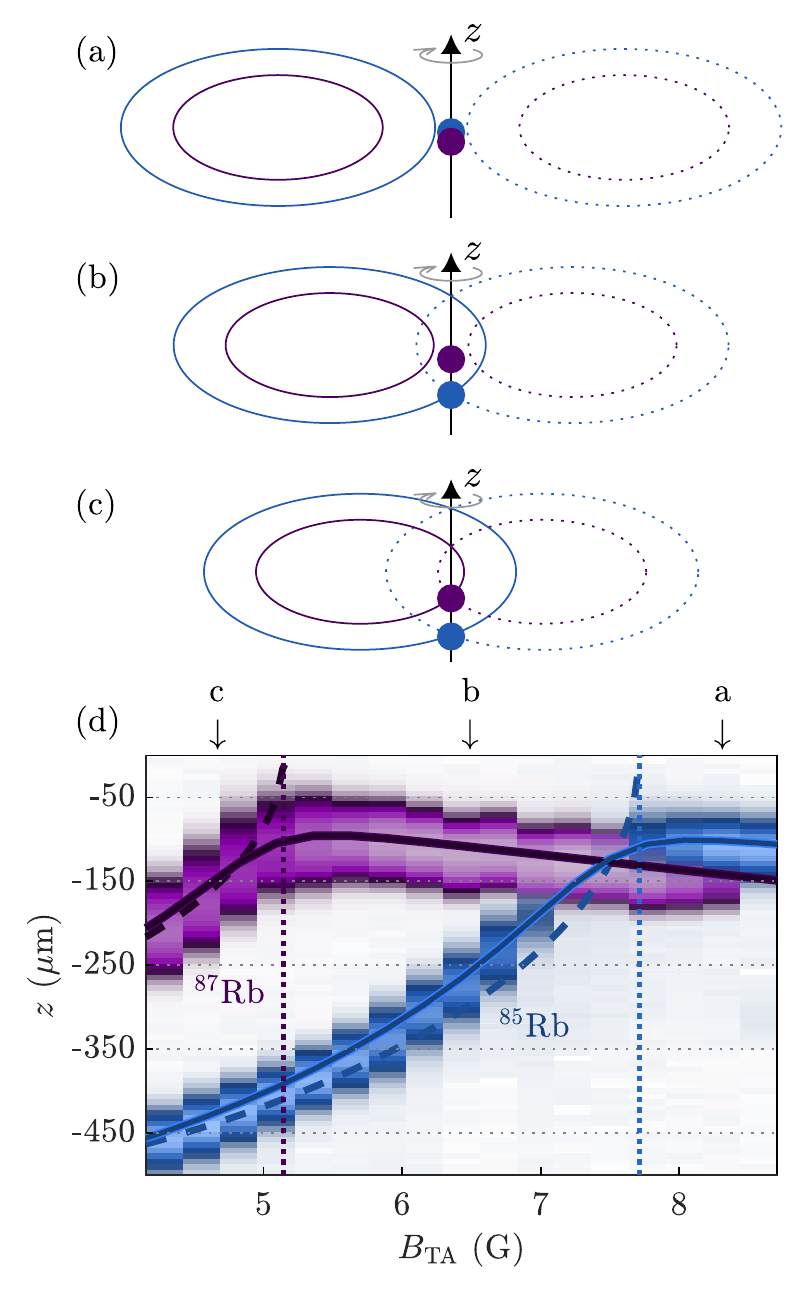}
		\caption{
			\label{fig:SSTAAP_Positions}
			(a-c) The different operating regimes of the dual-species TAAP.
			Filled circles show the locations of potential minima for \Rb{85} (blue) and \Rb{87} (purple). The ellipses show the resonant spheroids at phases $\omega_\mathrm{TA}t=0$ (solid lines) and $\pi$ (dotted lines) of the rotating field \Btop*{}.
			Three distinct regimes are shown: (a) $\hbar \omega < \gF{85} \muB{} \Btop{} $, (b) $\gF{85}\muB{} \Btop{} < \hbar \omega < \gF{87} \muB{} \Btop{}$, (c) $\gF{87} \muB{} \Btop{} < \hbar \omega$.
			(d) Measurements of the vertical position of \Rb{85} and \Rb{87} clouds as a function of \Btop{}.
			The observed density distribution for each species along the vertical direction is shown as a vertical slice for each unique value of \Btop{}.
			The dotted vertical lines show the two \rf{} resonances, where $\gF{i}\muB{}\Btop{} = \hbar \omega$.
			For each species, the colored lines show the value of $z$ from Eq.\ \eqref{eq:TAAPminpos} (dashed), and the numerically calculated TAAP trap minimum (solid).
		}
	\end{center}
	\end{figure}
	
	The TAAP differs for species with $g$-factors of different magnitude, such as \Rb{85} and \Rb{87} in their lower hyperfine states, which have $\gF{85}\!=\!-1/3$ and $\gF{87}\!=\!-1/2$.
	\refform{fig:SSTAAP_Positions} shows how the positions of each species change as a function of the time-averaging field \Btop{}.
	With $\Btop{} > \hbar \omega / \gF{85} \muB{}$, the resonant spheroids for both species orbit far from the atoms, which are confined near the origin.
	This scheme is illustrated in \refform{fig:SSTAAP_Positions}{a}.
	For $\gF{85} \muB{} \Btop{} < \hbar \omega < \gF{87} \muB{} \Btop{}$,
	only the resonant spheroid for \Rb{85} intersects the rotation axis, confining \Rb{85} in the lower well of the TAAP but keeping \Rb{87} confined near the origin by the TOP-like trap, as in \refform{fig:SSTAAP_Positions}{b}.
	In this configuration, the vertical position of the \Rb{85} potential minimum is strongly affected by \Btop{}, while that of \Rb{87} is not.
	When $\Btop{} < \hbar \omega/\gF{87} \muB{}$, the resonant spheroids for both species intersect the rotation axis, and both are loaded into the lower well of their respective TAAPs, as in \refform{fig:SSTAAP_Positions}{c}.
	
	\subsection{Measuring inelastic loss}
	
	To observe inelastic loss between \Rb{85} and \Rb{87}, we work in the regime in which the clouds of the two species are spatially overlapped, which requires that $\gF{85} \muB{} \Btop > \hbar \omega$, as shown in \refform{fig:SSTAAP_Positions}.
	The two isotopes are held in contact for a specified duration, then the remaining atom numbers \AN{i} of both are measured using absorption imaging.
	The raw images are processed using the fringe-removal algorithm developed by Ockeloen \etal{}~\cite{Ockeloen2010}.
	The temperatures $\Temperature{\mathit{i}}$ of both species are measured using time-of-flight expansion.
	
	The mixtures used in this work have atom numbers \AN{85}, \AN{87} of the two species, with $\AN{85} \ll \AN{87}$;
	the decrease in \AN{85} over time provides a clear signal to measure the inelastic loss rate.
	The fractional decrease in \AN{87} is negligible and cannot be distinguished above shot-to-shot variations.
	The inelastic collisions have a negligible effect on the temperature of \Rb{87}, and the \Rb{87} atoms thus provide a large bath of nearly constant density \density{87}.
	
	Our \rf{}-dressed trap confines two states of \Rb{85}, with $\dressedState\!=\!1, 2$.
	The two separate clouds that correspond to these states are discernible in absorption images, but their overlap means that only the total atom number $\AN{85}(t) = \AN{1} + \AN{2}$ can be measured accurately.
	For our experiments, initially $\AN{2} \gg \AN{1}$ because the method used to load the \rf{}-dressed trap favours projection from $\hyperfinem{}=-2$ into $\dressedState{}=2$.
	
	\section{Inelastic Loss}	
	\label{sec:Results}
	
	Including up to 3-body collision processes, $\AN{85}(t)$ decreases at a rate given by
	\begin{align}
	\label{eq:85Loss}
	\frac{d \AN{85}(t)}{d t} = & - \frac{\AN{85}(t)}{\tau_{85}} \nonumber \\
	& - \int \density{85}(t) \left(\kTwo{85,87} \density{87}(t) + \kThree{85,87,87} \density{87}(t)^2 \right) dV \nonumber \\
	& - \int \density{85}^2(t) \left(\kTwo{85,85} + \kThree{85,85,87} \density{87}(t) \right) dV \nonumber \\
	& - \int \density{85}^3(t) \left(\kThree{85,85,85}\right) dV,
	\end{align}
	where $\density{i}(t)$ are atom number densities, and $\tau_{85}$ is the lifetime of trapped \Rb{85} atoms from one-body losses and collisions with the background gas.
	The coefficients $k_j$ are $j$-body rate coefficients, with the colliding species indicated by the superscript.
	
	For pure \Rb{85} samples, \refform{eq:85Loss} reduces to
	\begin{equation}
	\label{eq:85OnlyLoss}
	\frac{d \AN{85}(t)}{d t} = -\frac{\AN{85}}{\tau_{85}} - \int \kTwo{85,85} \density{85}^2 dV - \int \kThree{85,85,85} \density{85}^3 dV.
	\end{equation}
	When only \Rb{85} is present, we observe an exponential decay of $\AN{85}(t)$ with lifetime $\tau_{85} = \SI{43}{\second}$, and the trapped atoms heating at a rate of \SI{74}{\nano\K\per\second} from an initial temperature of \SI{1}{\micro\K}.
	The fitted rate coefficients \kTwo{85,85} and \kThree{85,85,85} are consistent with zero, with upper bounds of $\kTwo{85,85}<\SI{3e-12}{\centi\m\cubed\per\second}$ and $\kThree{85,85,85}<\SI{e-22}{\centi\m\tothe{6}\per\second}$ in the \rf{}-dressed trap.
	These bounds are sufficiently low that the intraspecies inelastic loss is negligible for all experiments discussed in this work.
	
	When both species are present, inelastic collisions cause a rapid loss of \Rb{85}, with almost all atoms lost after a few hundred milliseconds.
	We argue that the loss occurs through two-body \Rb{87}+\Rb{85} collisions, as follows.
	Neglecting both the intraspecies loss and one-body loss, \refform{eq:85Loss} approximates to
	\begin{align}
	\label{eq:85LossWith87}
	\frac{d\AN{85}(t)}{d t} \approx & - \int n_{85}(t) \left(\kTwo{85,87} n_{87} + \kThree{85,87,87} n_{87}^2\right) dV \nonumber \\
	& - \int \kThree{85,85,87} n_{85}^2(t) n_{87} dV.
	\end{align}
	Thus, depending on which term dominates, $d\AN{85}/dt$ is proportional to either \density{85} or $\density{85}^2$.
	For an atomic cloud at constant temperature, $\density{i} \propto \AN{i}$.
	
\begin{figure}
		\begin{center}
			\centering{}
			\includegraphics[width=0.5\textwidth]{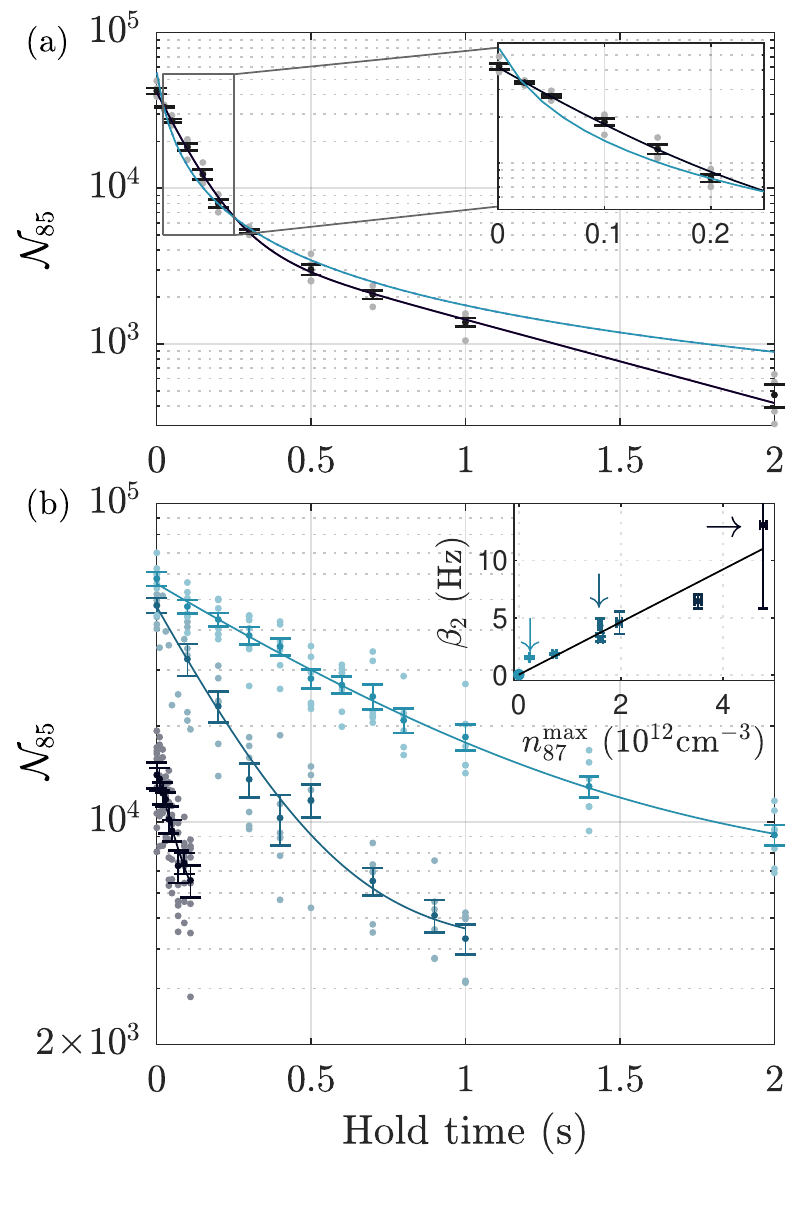}
			\caption{
				\label{fig:85Decay_FitModel}
				(a) The measured total \Rb{85} atom number \AN{85} as a function of hold time in the trap.
				The solid black line shows the best fit of a model in which the population of both trapped states of \Rb{85} decays exponentially.
				The solid blue line shows an alternative model in which each population decays with a rate proportional to $\AN{i}^2$.
				(b) At short hold times the change in atom number is dominated by the exponential atom loss from the state with $\dressedState\!=\!2$.
				The decay rates in the presence of three different atom number densities of \Rb{87} are shown.
				Inset: The fitted rate coefficients $\beta_2$ are linearly proportional to the peak \Rb{87} atom number density, $n_{87}^\text{max}$. Arrows indicate the three sets that are plotted in the outer panel.
			}
		\end{center}
	\end{figure}

	In \refform{fig:85Decay_FitModel}a, we show measurements of \AN{85} against hold time.
	The total \Rb{85} atom number is well described by a model where the two trapped states of \Rb{85} each decay exponentially, $\AN{85} = \AN{1} + \AN{2}$ with $\AN{i} = A_i e^{-\beta_i t}$.
	The different decay constants $\beta_i$ arise from the differing overlap of each state's density distribution with that of \Rb{87}.
	For the measurements in \refform{fig:85Decay_FitModel} the overlap of \Rb{85} atoms in the $\dressedState{}\!=\!2$ state with \Rb{87} is optimized, thus $\beta_2 \gg \beta_1$.
	A model in which $d\AN{i}(t)/dt \propto \AN{i}^2$ shows poor agreement with the data, as shown in \refform{fig:85Decay_FitModel}a.
	From $d\AN{i}/dt \propto \AN{i}$, it follows that one \Rb{85} atom is involved in each inelastic collision.
	
	For short hold times, $t \ll 1/\beta_1$, the atom number decays exponentially as 
	\begin{equation}
	\AN{85}(t) \approx A_{2} e^{-\beta_2 t} + A_{1}.
	\label{eq:N85ExpDecay}
	\end{equation}
	\refform{fig:85Decay_FitModel}b shows \AN{85} as a function of time during the initial fast exponential decay, for different densities of \Rb{87}.
	We fit \refform{eq:N85ExpDecay} to the data in \refform{fig:85Decay_FitModel}b, and in the inset plot $\beta_2$ against $n_{87}^\text{max}$, the maximum atom number density of \Rb{87}, which occurs at the centre of the trap.
	The measured decay rate is proportional to $n_{87}^\text{max}$, indicating that the inelastic collisions involve a single \Rb{87} atom.
	Thus we deduce that the inelastic loss arises from a mechanism involving two-body \Rb{87}+\Rb{85} collisions.
	
	\subsection*{Measuring the two-body rate coefficient}\label{sec:measuring-the-rate-constant}
	
	Having determined that two-body \Rb{87}+\Rb{85} inelastic collisions are the dominant loss mechanism for \Rb{85} in the trapped mixture, we now measure the two-body rate coefficient.
	\refform{eq:85LossWith87} further approximates to
	\begin{equation}
	\frac{d \AN{85}}{d t} \approx - \int \kTwo{85,87} n_{85} n_{87} dV
	\end{equation}
	We measure the inelastic loss rate by fitting \refform{eq:N85ExpDecay} to $\AN{85}(t)$.
	As only the total inelastic loss rate is measurable, and \kTwo{85,87} varies with position in the trap, we are able to extract only a mean value of \kTwo{85,87} that is weighted by the overlap between the species,
	\begin{equation}
	\langle \kTwo{85,87} \rangle = \frac{\int \kTwo{85,87} n_{85} n_{87} dV}{\int n_{85} n_{87} dV},
	\end{equation}
	where the term $\int n_{85} n_{87} dV$ is the overlap integral that quantifies the spatial overlap. Hence,
	\begin{equation}
	\frac{d\AN{85}}{dt} \approx - \langle \kTwo{85,87} \rangle \int n_{85} n_{87} dV.
	\end{equation}	
	
	Determining the overlap integral requires knowledge of the atom number densities $n_{i}$.
	We calculate these densities using measured values of the cloud temperatures, quadrupole field gradient, rotating bias field amplitude, and \rf{} field.
	The temperatures are determined from time-of-flight expansion of the clouds, and we find that \Temperature{87} is independent of hold time.
	However, it is not possible to determine \Temperature{85} at arbitrary hold times; the significant \Rb{85} atom loss results in weak absorption imaging signals, which cannot be reliably fitted with Gaussian profiles.
	Instead, we determine \Temperature{85} at $t=0$ and assume it is constant thereafter.
	A Monte-Carlo method is used to determine the uncertainties in $\int \density{85} \density{87} dV$, which incorporate the individual uncertainties (including systematic errors) of all independent parameters.
	The uncertainties in $\int \density{85} \density{87} dV$ are combined in quadrature with those of the fitted decay rates to determine the uncertainty of $\langle \kTwo{85,87} \rangle$.
	
\begin{figure}
		\begin{center}
			\centering{}
			\includegraphics[width=0.45\textwidth]{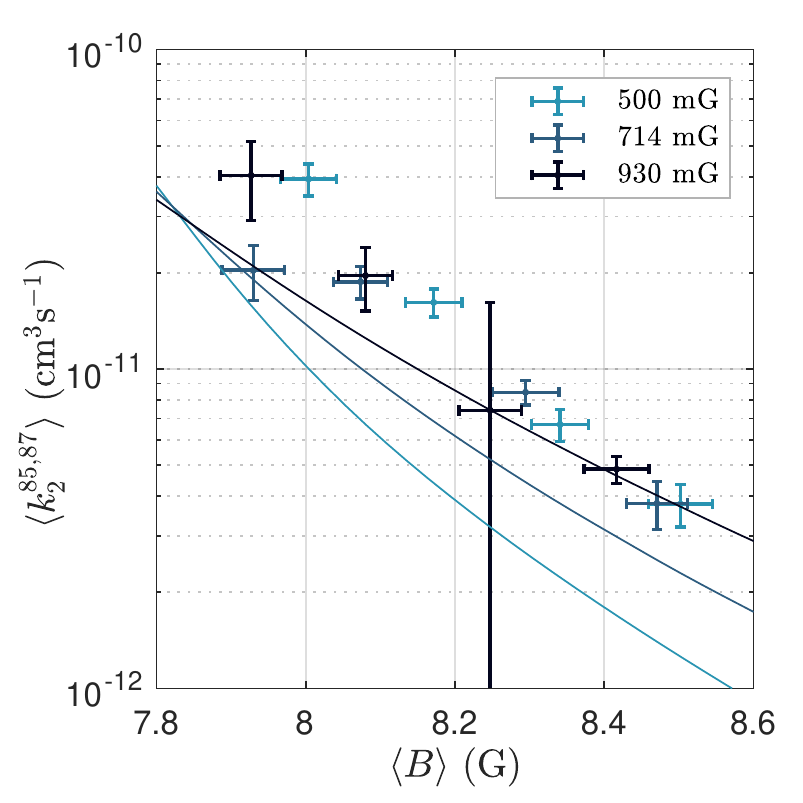}
			\caption{
				\label{fig:k2vB}
				Measurements of $\langle \kTwo{85,87} \rangle$ as a function of the average magnetic field $\langle \Bzero{} \rangle$ for three amplitudes \Brf{} of a \SI{3.6}{\mega\Hz} \rf{}-dressing field.
				The solid lines show values of $\langle \kTwo{85,87} \rangle$ for each \rf{} amplitude as predicted from coupled-channel calculations.
				}
		\end{center}
	\end{figure}

	We explore the dependence of $\langle \kTwo{85,87} \rangle$ on the static magnetic field by adjusting \Btop{}, which is akin to a bias field in our setup.
	This is possible provided the two species remain overlapped, which requires that $\Btop{} > \hbar \omega /(\gF{85}\muB{})$, as described in \refform{sec:SSTAAP}.
	For any given value of \Btop{}, collisions occur over a range of different static magnetic fields because of the field gradient that is required to confine the atoms.
	As such, we compare our measured rate coefficients as a function of the overlap-weighted average $\langle \Bzero \rangle$, defined analogously to $\langle \kTwo{85,87} \rangle$.
	Our measurements are shown in \refform{fig:k2vB}, for three different amplitudes of the \rf{}-dressing field.
	We observe that the two-body rate coefficient increases with decreasing $\langle \Bzero \rangle$, and within the uncertainties observe no clear dependence on \rf{} amplitude.
	We also plot the values of $\langle\kTwo{85,87}\rangle$ predicted from our scattering calculations, which are described in the next section.
	
	\section{Quantum scattering calculations}
	\label{sec:QuantumScatteringCalculations}
	
	We model the collisional losses by carrying out quantum-mechanical scattering
	calculations using the MOLSCAT program~\cite{molscat2019,mbf-github:2019}.
	The method used was described in ref.\ \onlinecite{Owens2017} for \rf{} polarization in the plane perpendicular to the magnetic field, and is summarized in \refform{appendix:QSC_NumericalMethods}.
	It has antecedents in refs.\ \onlinecite{Kaufman2009,Tscherbul2010,Hanna2010}.
	The wave function for a colliding pair of atoms is expanded in an uncoupled \rf{}-dressed basis set,
	\begin{equation}
	|s_1 m_{s1}\rangle |i_1 m_{i1}\rangle |s_2 m_{s2}\rangle |i_2 m_{i2}\rangle |L
	M_L\rangle |N M_N\rangle, \label{eq:bas-u}
	\end{equation}
	where the indices ($1$, $2$) label quantities associated with the first and second atoms,
	$L$ is the angular momentum for relative motion of the two atoms, and $M_L$ is its projection onto the $z$ axis.
		
	\begin{figure*}[t]
	\includegraphics[width=\textwidth]{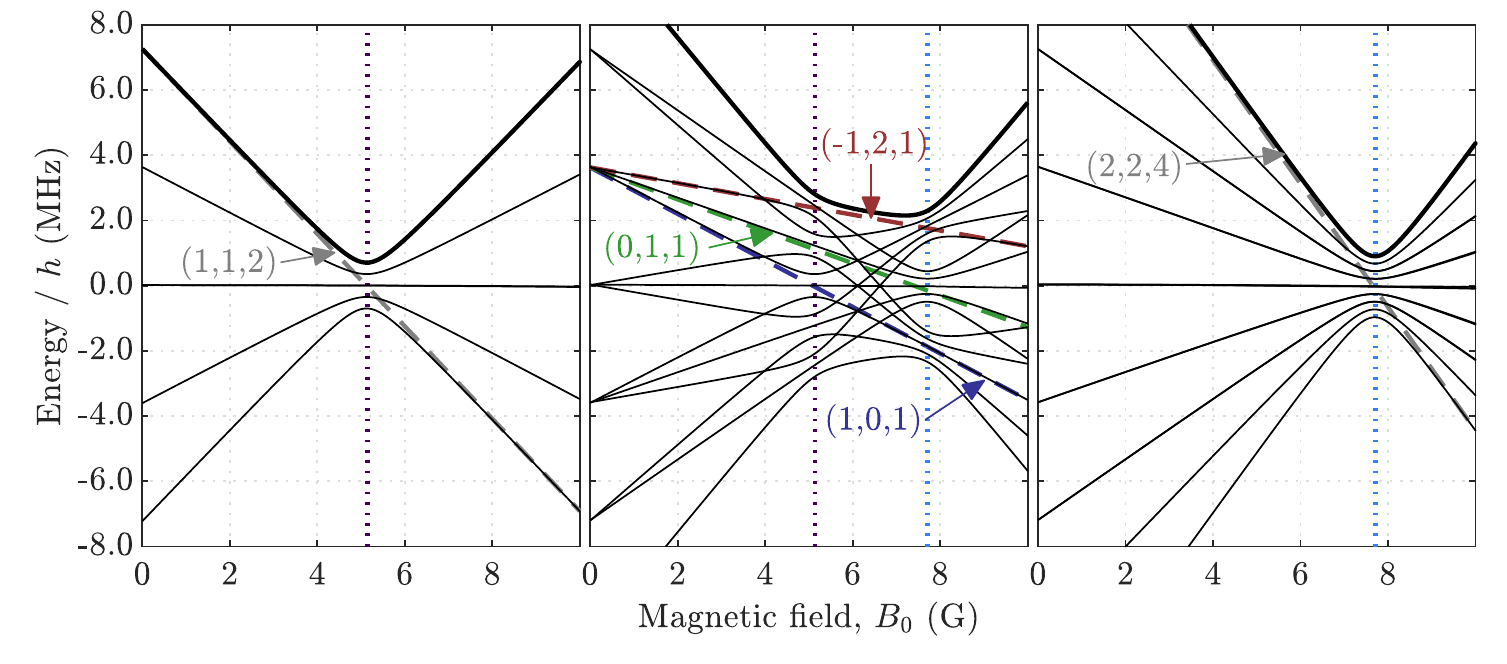}
	\caption{The \rf{}-dressed atomic thresholds (black, solid) for \Rb{87}+\Rb{87} (left),	\Rb{87}+\Rb{85} (center) and \Rb{85}+\Rb{85} (right) for $m_{f1}+m_{f2}+M_N=0$, with $\Brf{}=\SI{0.5}{\Gauss}$ at a frequency of \SI{3.6}{\mega\Hz}.
		Both atoms are in their lower hyperfine state.
		Selected thresholds are also shown for zero \rf{} intensity (dashed lines), labelled
		with quantum numbers $(m_{f1},m_{f2},N)$.
		The \rf{} resonances for each species are indicated by vertical dotted lines.
		The thresholds corresponding to collisions of trapped atoms in this work are indicated in bold.
		}
	\label{fig:Rb2thresh}
	\end{figure*}
	
	To understand collisions between trapped atoms, it is useful to consider the
	thresholds (i.e., the energies of separated atomic pairs) as a function of
	magnetic field. Figure \ref{fig:Rb2thresh} compares the thresholds for
	\Rb{87}+\Rb{85} with those for \Rb{87}+\Rb{87} and \Rb{85}+\Rb{85}
	for an \rf{} field strength $\Brf{}=\SI{0.5}{\Gauss}$. Only states with $m_{f1}+m_{f2}+M_N=0$
	are shown; as discussed below, this quantity is conserved in spin-exchange
	collisions (though not in spin-relaxation collisions).
	
	\subsection{Homonuclear systems}
	
	The thresholds for the homonuclear systems are shown in the end panels of \refform{fig:Rb2thresh}.
	They show simple maxima or minima at a single magnetic field, $B=\SI{5.12}{\Gauss}$ for \Rb{87} and $B=7.70$~G for \Rb{85}.
	In all cases the trapped states correspond to the uppermost threshold of those shown, though other thresholds exist for different photon numbers or higher hyperfine states.
	For \Rb{87}+\Rb{87}, the uppermost state has character $(m_{f1},m_{f2},N)=(1,1,2)$ at fields below the crossing and $(-1,-1,-2)$ above it.
	For \Rb{85}+\Rb{85}, the uppermost state has character $(2,2,4)$ below
the crossing and $(-2,-2,-4)$ above it.

In both homonuclear cases, there are no lower-energy states in the same multiplet with the same photon number, so collisional decay can occur in only two ways~\cite{Owens2017}:

\noindent (1) Close to the crossing, the $m_f$ quantum numbers are mixed by the
		photon couplings, so that \rf{}-induced spin-exchange collisions can
		transfer atoms to lower thresholds without changing $L$ from 0.

\noindent (2) At fields above the crossing, the $m_f\!=\!-1$ state for \Rb{87} (or $m_f\!=\!-2$ state for \Rb{85}) is not the ground state.
		Even in the absence of \rf{}
		radiation, two $m_f=-1$ or $-2$ atoms can undergo spin-relaxation
		collisions that change both $M_F=m_{f1}+m_{f2}$ and $M_L$ (and thus must
		change $L$ from 0 to 2) but conserve $M_F+M_L$. Spin relaxation is
usually very slow, both because the spin-dipolar coupling $\hat V^{\rm d}(R)$
is very weak and because there is a centrifugal barrier higher than the kinetic
energy in the outgoing channel with $L=2$.

\Rb{87}+\Rb{87} is a special case, with very similar singlet and triplet
scattering lengths $a_{\rm s}=\SI{90.6}{\bohr}$ and $a_{\rm t}=\SI{98.96}{\bohr}$. This is known to suppress
spin-exchange collisions dramatically \cite{Myatt1997,Julienne1997,Burke1997}, and ref.\
\onlinecite{Owens2017} showed that it also suppresses \rf{}-induced spin-exchange
collisions. Thus collisional losses in \rf{}-dressed traps for pure \Rb{87} are
dominated by spin relaxation, somewhat modified by the \rf{} radiation
\cite{Owens2017}. \Rb{85}+\Rb{85} has $a_{\rm s}=\SI{2735}{\bohr}$ and $a_{\rm
t}=\SI{-386}{\bohr}$ \cite{Blackley2013}; although superficially very
different, these give similar values of the low-energy s-wave scattering
phase. As a result, \rf{}-free and \rf{}-induced spin-exchange collisions are
suppressed for pure \Rb{85} as well, though not as strongly as for \Rb{87}.
 	
	\subsection{Heteronuclear systems}
	
	The thresholds for \Rb{87}+\Rb{85} are very different from those for the
	homonuclear systems. The uppermost state has character
	$(m_{f1},m_{f2},N)=(1,2,3)$ at fields below the \Rb{87} resonance at \SI{5.12}{\Gauss}, and
	$(-1,-2,-3)$ above the \Rb{85} resonance at \SI{7.70}{\Gauss}. \rf{}-induced spin exchange is
	possible close to the crossings and \rf{}-modified spin relaxation is possible
	above \SI{5.12}{\Gauss}, as for the homonuclear systems.
	However, at magnetic fields	\emph{between} the two crossings the uppermost state has predominantly $(-1,2,1)$ character, and there are lower pair states that have predominantly $(0,1,1)$ and (1,0,1) character, with the same photon number and value of $M_F$, as shown by dashed lines in \refform{fig:Rb2thresh}.
	Spin-exchange collisions that transfer atoms to these lower thresholds are thus allowed in this intermediate region, even without the couplings due to \rf{} radiation.
	The scattering lengths for \Rb{87}+\Rb{85} are $a_{\rm s}=\SI{202}{\bohr}$ and $a_{\rm t}=\SI{12}{\bohr}$, so spin exchange is \emph{not} suppressed	in the mixture and fast losses are expected at these intermediate fields.
	
	\begin{figure}[t]
		\includegraphics[width=\columnwidth]{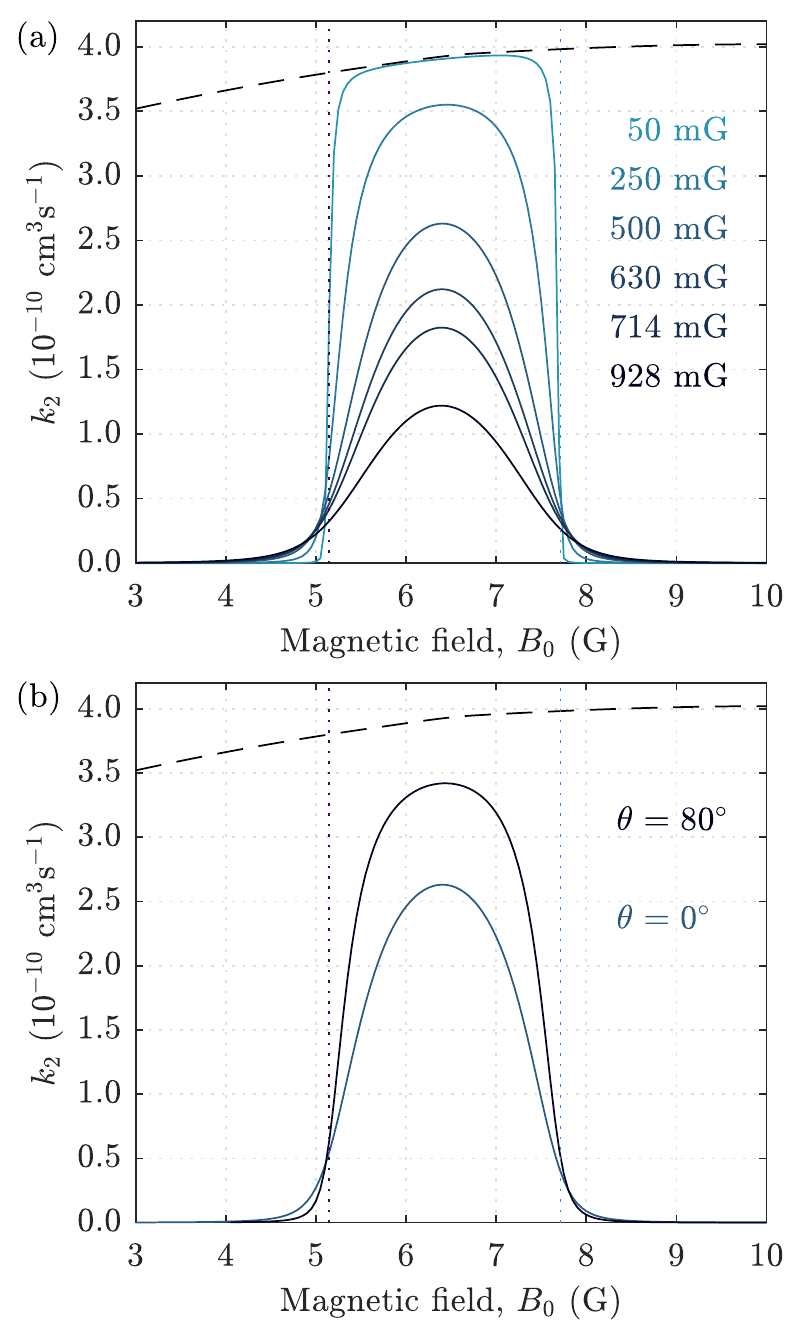}
		\caption{
			Rate coefficients for inelastic collisions between \rf{}-dressed \Rb{87} and \Rb{85} atoms in their hyperfine ground states, as a function of magnetic field, at a collision energy of \SI{0.4}{\micro\K}. 
			Results are shown for \rf{} radiation with $\sigma_-$ polarization at a frequency of \SI{3.6}{\mega\Hz}.
			(a) Dependence on \Brf{} with \rf{} polarization in the plane perpendicular to \Bzero{}.
			The dashed black line shows the rate coefficient for \rf{}-free spin exchange for $(\hyperfineState{},\hyperfinem{}) = \Rb{87}\,(1,-1) + \Rb{85}\,(2,2)$. The dotted lines indicate the magnetic field at which the \rf{} is resonant for \Rb{85} and \Rb{87}.
			(b) Dependence on the tilt angle \tiltAngle{}, for \Brf{} fixed at \SI{0.5}{\Gauss}. 
		} \label{fig:8785loss}
	\end{figure}
	
	\subsection{Calculated rates and comparison}
	
	\refform{fig:8785loss}{a} shows the calculated inelastic rate coefficients for collisions between \rf{}-dressed \Rb{87} atoms in $\hyperfineState\!=\!1$, $\dressedState\!=\!1$ and \Rb{85} atoms in $\hyperfineState\!=\!2$, $\dressedState\!=\!2$ as a function of magnetic field, for several \rf{} field strengths.
	The calculations were carried out for $\sigma_-$ polarization in the plane perpendicular to \Bzero{}.
	In these calculations $L_{\rm max}=0$, so spin-relaxation collisions are excluded.
	At the lowest \rf{} field strength, $\Brf{}=\SI{50}{\milli\Gauss}$, the avoided crossings between the thresholds are very sharp and the states are well described by the quantum numbers $(\hyperfinem{1},\hyperfinem{2},N)$ introduced at the start of \refform{sec:QuantumScatteringCalculations}.
	In this regime the spin-exchange losses are forbidden below the \Rb{87} radiofrequency resonance at \SI{5.12}{\Gauss} and above the \Rb{85} radiofrequency resonance at \SI{7.70}{\Gauss}, but at intermediate fields they occur at almost the full \rf{}-free rate for $(f,m_f)=(1,-1)+(2,2)$ collisions, shown as the black dashed line.
	At higher \rf{} field strengths, the avoided crossings extend further into the intermediate field region; the uppermost state is a mixture of $(m_{f1},m_{f2},N)=(-1,2,1)$ and other pair states that do not decay as fast.
	The effect is to broaden the edges of the flat-topped peak that exists for $\Brf{}=\SI{0.05}{\Gauss}$ and depress the height of the peak in the central region.
	
	In the experiment, the atoms are trapped at locations where the magnetic
	field is not perpendicular to the plane of circular polarization.
	To explore the effects of this, we carried out additional calculations where the radiation is still $\sigma_-$ polarized in the plane perpendicular to $z$, but the static magnetic field $B_0$ is tilted by an angle $\theta$ from the $z$ axis.
	In this case $M_\textrm{tot}$ is
	no longer conserved, resulting in an increase in the number of open
	channels.
	For $L_\textrm{max} = 0$, the number of open channels increases from 15 to 56.
	The calculated loss profiles are shown in \refform{fig:8785loss}{b} for $\Brf{}=\SI{0.5}{\Gauss}$ and different values of the tilt angle $\theta$;
	the profile remains qualitatively similar to that at $\theta=0$, especially far from the avoided crossings in \refform{fig:Rb2thresh}, where the magnetic field dominates.
	However, the onset of loss is sharper for tilted fields, resembling that at smaller values of \Brf{} in \refform{fig:8785loss}{a}.
	
	At fields above the \rf{} resonance at \SI{5.12}{\Gauss}, spin relaxation can also
	occur. For example the state $(m_{f1},m_{f2},M_L,N)=(-1,2,0,1)$ can decay to
	$(0,2,-1,1)$, $(1,2,-2,1)$, $(-1,1,1,1)$ or $(-1,0,2,1)$, conserving $M_F+M_L$.
	This spin-relaxation loss is slower than the spin-exchange losses
	considered in this paper by five orders of magnitude and so is neglected
	in the analysis.
	
	Overlap-weighted averages $\langle \kTwo{85,87} \rangle$ of the calculated rate coefficients \kTwo{85,87} are plotted as solid lines alongside the experimental data in \refform{fig:k2vB}.
	To perform the overlap-weighted averaging, we calculate the spatial distributions \density{85}, \density{87} using the average temperature, atom number and trapping fields for each particular value of \Brf{} shown.
	We numerically integrate these density distributions to determine $\langle \kTwo{85,87} \rangle = \int \density{85} \density{87} \kTwo{85,87} dV$, taking into account the variation of \kTwo{85,87} with \Bzero{}.
	The tilt angle $\theta$ varies by only a few degrees across the region where the two species overlap, and so we use a constant $\theta=\ang{80}$ for the calculations.
	
	Our calculated values of $\langle \kTwo{85,87} \rangle$ are in reasonable agreement with the experimental measurements shown in \refform{fig:k2vB}.
	The measurements clearly demonstrate that $\langle\kTwo{85,87}\rangle$ increases with magnetic field as the \Rb{85} resonance is approached from the high-field side, which is consistent with the predicted rate coefficients.
	No clear trend with \rf{} amplitude is discernible in our measured data, although this would be difficult to observe given our uncertainties.
	In general, the measured rates are slightly higher than the predicted values.  
	This discrepancy could be caused by a systematic error that underestimates the atom numbers \AN{85} and \AN{87}.
	
	\section{Semiclassical interpretation}
	\label{sec:semiclassical}
	
	Low inelastic loss rates in \rf{}-dressed potentials were previously measured in experiments using \Rb{87}.
	Those results were interpreted using a semiclassical model~\cite{Hofferberth2006,GarrawayPerrinReview}, which we now revisit in light of our work.
	The model was first introduced in the context of microwave dressing~\cite{Agosta1989}, and it has also been applied to collisions during \rf{} evaporative cooling~\cite{Moerdijk1996}.
	
	Before we discuss the collision model, we first recap the semiclassical picture of an atom in an \rf{} field.
	The Hamiltonian of a single atom interacting with a magnetic field is
	\begin{equation}
	\Hamiltonian{} = \gF{} \muB{} \vec{\hyperfineState} \cdot \left(\Brf*{}(t) + \Bzero*{} \right).
	\label{eq:semiclassical_atomInField_Hamiltonian}
	\end{equation}
	The time-dependence is removed by transforming into a frame that rotates with the \rf{} field, with coordinate axes
	\begin{align}
	\label{eq:xz_axes_def}
	\coordaxis{x}{}' =& \ \coordaxis{x} \cos(\drf t) + \coordaxis{y} \sin(\drf t), \nonumber \\
	\coordaxis{z}{}' =& \ \coordaxis{z},
	\end{align}
	followed by making the rotating-wave approximation.
	The resulting time-independent Hamiltonian is
	\begin{equation}
	\Hamiltonian{RWA} = \hbar \left(\Omega \coordaxis{x}{}' + \delta \coordaxis{z}{}'\right) \cdot \vec{\hyperfineState},
	\label{eq:rwaHamiltonian}
	\end{equation}
	where $\delta$ is the angular frequency detuning and $\Omega$ the resonant Rabi frequency, defined previously.
	Diagonalising this semi-classical Hamiltonian gives the eigenenergies of an atom in the applied magnetic fields.
	
\begin{figure}
		\begin{center}
			\includegraphics[width=\columnwidth]{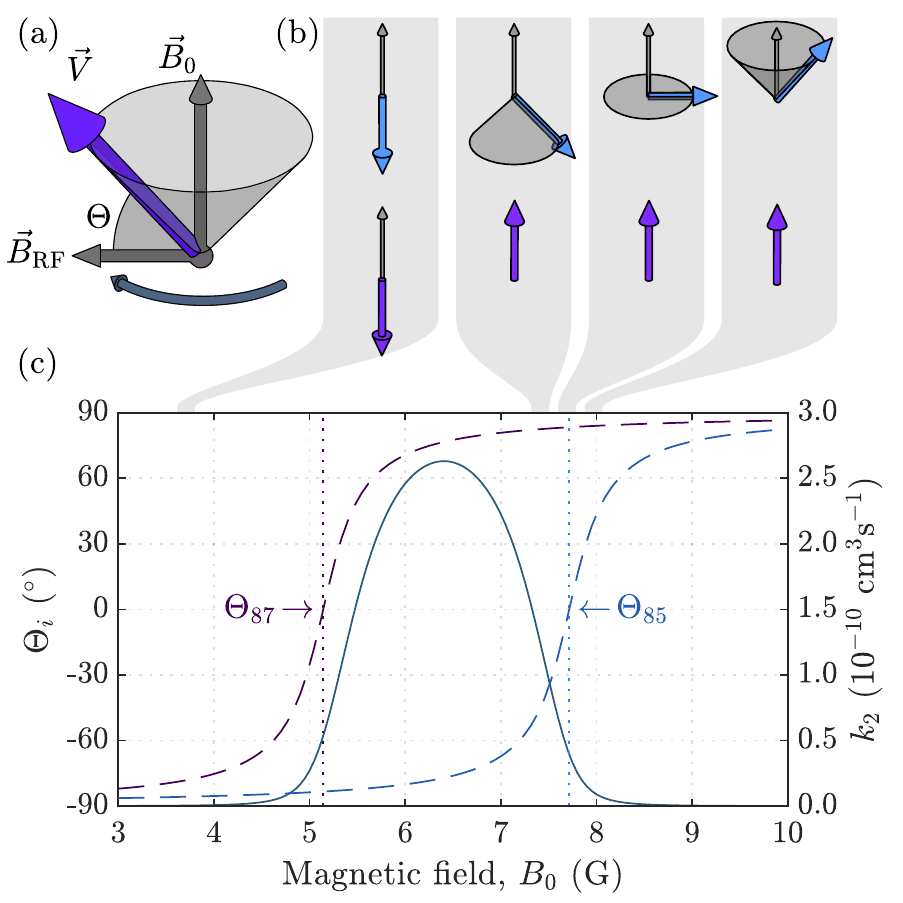}
		\end{center}
		\caption{The semiclassical picture.
			(a) \Beff{} precesses around \Bzero*{} at the \rf{} frequency, and is coplanar with both \Brf{} and \Bzero*{}. 
			The angle \scTiltAngle{} is defined in the text.
			(b) Illustrations of \Beff{} for \Rb{85} (top row) and \Rb{87} (bottom row).
			Each frame illustrates the fields at different magnetic fields \Bzero{}.
			(c) The angle \scTiltAngle{i} is shown for each species as a function of magnetic field for a \SI{3.6}{\mega\Hz} \rf{} dressing field.
			Dotted vertical lines mark the \rf{} resonances for each species.
			The rate coefficient \kTwo{85,87} for $\theta=0$, $\Brf{}=\SI{0.5}{\Gauss}$ is overlaid for comparison.
			\label{fig:semiclassical}
		}
	\end{figure}

	\Hamiltonian{RWA} is proportional to the dot product of $\vec{\hyperfineState{}}$ with the vector \Beff{},
	\begin{equation}
	\label{eq:Beff_def}
	\Beff{} = \left(\Omega \coordaxis{x}{}' + \delta \coordaxis{z}{}'\right).
	\end{equation}
	Consequently, the eigenstates of \Hamiltonian{RWA} have a well-defined projection $\dressedState$ of $\vec{\hyperfineState{}}$ in the direction of \Beff{}.
	\refform{fig:semiclassical}a shows \Beff{} observed from the laboratory frame, in which it precesses about the static field \Bzero*{} at the angular frequency \drf{} and with angle $\scTiltAngle{} = \arctan(\delta/\Omega)$.
	
	The semiclassical model of \rf{}-dressed collisions posits that spin exchange does not occur between identical atoms that are in eigenstates of extreme \dressedState{}~\cite{Moerdijk1996,Hofferberth2006,GarrawayPerrinReview}.
	When two such atoms collide, their total angular momentum also has a maximum projection along \Beff{}, with $\widetilde{M}=\widetilde{m}_{1} + \widetilde{m}_{2}$.
	In the semiclassical picture, there are no other open channels with the same value of $\widetilde{M}$, thus spin-exchange collisions are forbidden by violation of angular momentum conservation.
	We stress that these conclusions are incorrect;
	the semiclassical picture neglects couplings to the \rf{} field during collisions, and therefore fails to predict the \rf{}-induced spin exchange described earlier.
	The rate coefficient for \rf{}-induced spin exchange is usually large, but this is not the case for either \Rb{87} or \Rb{85};
	it appears that the low inelastic loss rates observed previously for \rf{}-dressed \Rb{87} atoms are in agreement with the semiclassical model's predictions only by coincidence.

	The same semiclassical model predicts that spin exchange can occur when two atoms with different values of $|\gF{}|$ collide, even if they are in states of maximum \dressedState{}.
	The vectors $\Beff{}_{85}$, $\Beff{}_{87}$ of the two species are in general not parallel, due to the different Rabi frequencies and detunings from the \rf{} resonance.
	They precess around \Bzero*{} at the same angular frequency $\omega$, but with different angles $\scTiltAngle{85}, \scTiltAngle{87}$. 
	These vectors are illustrated for different \Bzero{} in \refform{fig:semiclassical}b, with the associated angles $\scTiltAngle{85}$ and $\scTiltAngle{87}$ shown in \refform{fig:semiclassical}c.
	
	At fields much greater than \SI{7.70}{\Gauss}, the detunings of both species are large and positive.
	Both angles tend to $\pi/2$, and the vectors $\Beff{}_{85}$ and $\Beff{}_{87}$ are nearly parallel.
	In this case, a $\Rb{87}+\Rb{85}$ pair has an extreme value of the total angular momentum $\widetilde{M}$ when each atom is in an eigenstate of maximum $\dressedState{}$.	
	Spin exchange is forbidden on the grounds of angular momentum conservation, as for identical atoms, and the inelastic rate coefficient \kTwo{85,87} is small.
	A similar argument follows for very weak fields below \SI{5.12}{\Gauss}, where the detunings are large and negative and both \scTiltAngle{85} and \scTiltAngle{87} tend to $-\pi/2$.
	
	At intermediate fields, the angles \scTiltAngle{85} and \scTiltAngle{87} are very dissimilar.
	The vectors $\Beff{}_{85}$, $\Beff{}_{87}$ are misaligned, and spin exchange is \emph{not} forbidden on grounds of angular momentum conservation.
	The rate coefficient increases as the angles diverge, and peaks at the midpoint between the two \rf{} resonances, where $\Beff{}_{85}$, $\Beff{}_{87}$ are almost antiparallel to each other.
	The \rf{} amplitude determines how slowly \scTiltAngle{85} and \scTiltAngle{87} change with respect to magnetic field in the vicinity of the \rf{} resonances, and larger \rf{} amplitudes therefore broaden the edges of the \kTwo{85,87} peak over a wider range of magnetic fields.
	
	Finally, we remark that the semiclassical model predicts that spin exchange occurs for collisions between atoms with $g$-factors of different sign, even when the magnitudes of the $g$-factors are the same;
	although angles \scTiltAngle{\hyperfineState=1} and \scTiltAngle{\hyperfineState=2} are matched, the \Beff{}s precesses with different handedness around the static field for each species.
	Further work is required to compare the semiclassical and quantal pictures with experimental data of different hyperfine states.
	
	\section{Conclusion}
	\label{sec:Conclusion}
	
	\begin{table}
		\begin{center}
			\begin{tabular}{lrrr}
				\toprule{}
				Mixture & $a_\mathrm{s}$ (\si{\bohr}) & $a_\mathrm{t}$ (\si{\bohr}) & Ref. \\ \colrule{}
				& & & \\[-1em]
				\Lithium{6}+\Lithium{6} & $45.154$ & $-2113$ & \cite{Zurn2013} \\
				\Lithium{7}+\Lithium{7} & $34.331(2)$ & $-26.92(7)$ & \cite{Julienne2014} \\
				& & & \\
				\Lithium{6}+\Sodium{23} & $-73(8)$ & $-76(5)$ & \cite{Schuster2012} \\
				& & & \\
				\Lithium{7}+\Potassium{39} & $29.83$ & $81.99$ & \cite{Tiemann2009} \\
				\Lithium{7}+\Potassium{40} & $14.88$ & $75.27$ & \cite{Tiemann2009} \\
				\Lithium{7}+\Potassium{41} & $-6.375$ & $69.76$ & \cite{Tiemann2009} \\
				\Lithium{6}+\Potassium{39} & $64.93$ & $68.59$ & \cite{Tiemann2009} \\
				\Lithium{6}+\Potassium{40} & $52.61$ & $64.41$ & \cite{Tiemann2009} \\
				\Lithium{6}+\Potassium{41} & $42.75$ & $60.77$ & \cite{Tiemann2009} \\
				& & & \\
				\Lithium{7}+\Rb{87} & $54.75(30)$ & $-66.66(10)$ & \cite{PhysRevLett.115.043201} \\
				& & & \\
				\Lithium{6}+\Caesium{133} & $30.252(100)$ & $-34.259(200)$ & \cite{Repp2013} \\
				\Lithium{7}+\Caesium{133} & $45.477(150)$ & $908.6(100)$ & \cite{Repp2013} \\
				& & & \\
				\Sodium{23}+\Sodium{23} & $18.81(80)$ & $64.30(40)$ & \cite{Knoop2011} \\ 
				& & & \\
				\Sodium{23}+\Potassium{39} & $311.8(20)$ & $-83.97(50)$ & \cite{Zhu2017} \\
				\Sodium{23}+\Potassium{40} & $66.7(3)$ & $-824.7(30)$ & \cite{Zhu2017} \\
				\Sodium{23}+\Potassium{41} & $3.39(20)$ & $267.05(50)$ & \cite{Zhu2017} \\
				& & & \\
				\Sodium{23}+\Rb{87} & $109$ & $70$ & \cite{Pashov2005} \\
				\Sodium{23}+\Rb{85} & $396$ & $81$ & \cite{Pashov2005} \\
				& & & \\
				\Sodium{23}+\Caesium{133} & $428(9)$ & $30.4(0.6)$ & \cite{Hood2019} \\
				& & &  \\
				\Potassium{39}+\Potassium{39} & $138.49(12)$ & $-33.48(18)$ & \cite{Falke2008} \\
				\Potassium{39}+\Potassium{40} & $-2.84(10)$ & $-1985(69)$ & \cite{Falke2008} \\
				\Potassium{39}+\Potassium{41} & $113.07(12)$ & $177.10(27)$ & \cite{Falke2008} \\
				\Potassium{40}+\Potassium{40} & $104.41(9)$ & $169.67(24)$ & \cite{Falke2008} \\
				\Potassium{40}+\Potassium{41} & $-54.28(21)$ & $97.39(9)$ & \cite{Falke2008} \\
				\Potassium{41}+\Potassium{41} & $85.53(6)$ & $60.54(6)$ & \cite{Falke2008} \\
				& & & \\
				\Potassium{39}+\Rb{85} & $26.5(0.9)$ & $63.0(0.5)$ & \cite{Ferlaino2006} \\
				\Potassium{39}+\Rb{87} & $824^{+90}_{-70}$ & $35.9(0.7)$ & \cite{Ferlaino2006} \\
				\Potassium{40}+\Rb{85} & $64.5(0.6)$ & $-28.4(1.6)$ & \cite{Ferlaino2006} \\
				\Potassium{40}+\Rb{87} & $-111(5)$ & $-215(10)$ & \cite{Ferlaino2006} \\
				\Potassium{41}+\Rb{85} & $106.0(0.8)$ & $348(10)$ & \cite{Ferlaino2006} \\
				\Potassium{41}+\Rb{87} & $14.0(1.1)$ & $163.7(1.6)$ & \cite{Ferlaino2006} \\
				& & & \\
				\Potassium{39}+\Caesium{133} & $-18.37$ & $74.88(9)$ & \cite{Grobner2017} \\
				\Potassium{40}+\Caesium{133} & $-51.44$ & $-71.67(45)$ & \cite{Grobner2017} \\
				\Potassium{41}+\Caesium{133} & $-72.79$ & $179.06(28)$ & \cite{Grobner2017} \\
				& & &  \\
				\Rb{85}+\Rb{85} & $2735$ & $-386$ & \cite{Blackley2013} \\
				\Rb{85}+\Rb{87} & $202$ & $12$ & present work \\
				\Rb{87}+\Rb{87} & $90.6$ & $98.96$ & \cite{Marte2002} \\
				& & & \\
				\Rb{85}+\Caesium{133} & $585.6$ & $11.27$ & \cite{Cho2013} \\
				\Rb{87}+\Caesium{133} & $997(11)$ & $513.3(2.2)$ & \cite{Takekoshi2012} \\
				& & & \\
				\Caesium{133}+\Caesium{133} & $286.5(1)$ & $2858(19)$ & \cite{Berninger2013} \\
				& & & \\
				\botrule{}
			\end{tabular}
		\end{center}
	\caption{\label{tab:ScattLengths}Singlet $a_\mathrm{s}$ and triplet $a_\mathrm{t}$ scattering lengths for isotopic mixtures of alkali-metal atoms. The uncertainties have been added where possible.}
	\end{table}
	
	In this paper we have investigated the inelastic collisions that occur in an \rf{}-dressed mixture of \Rb{85} and \Rb{87}.
	We measured the loss of a small population of \Rb{85} atoms in the presence of a larger \Rb{87} bath, and identified the dominant mechanism as two-body \Rb{87}+\Rb{85} inelastic collisions.
	The inelastic rate coefficient \kTwo{85,87} was shown to vary as a function of magnetic field, with \kTwo{85,87} increasing as the atomic \rf{} resonance was approached from the high-field side.
	We used a theoretical model of \rf{}-dressed collisions to predict values of \kTwo{85,87}, and find they are in reasonable agreement with the measured values given that no free parameters were used to fit.
	
	When \rf{}-dressed potentials are used to confine atoms, the atoms are in states with a potential energy minimum at the atomic \rf{} resonance.
	If two atoms have different magnitudes of \gF{}, they are resonant with an applied \rf{} field at different values of the static field.
	At fields between these two values, the atoms are predominantly in states where spin-exchange collisions are allowed, even in the absence of coupling to the \rf{} field.
	Unless the singlet and triplet scattering lengths are similar, or the magnitudes of both are large, this spin exchange is expected to be fast.
	This contrasts with the situation when two atoms have very similar values of \gF{}, and are thus resonant at the same value of the static field.
	In this case spin-exchange collisions are forbidden except close to the trap center, where mixing of the Zeeman states by photon interactions permits \rf{}-induced spin exchange.
	This \rf{}-induced spin exchange can also be moderately fast unless the singlet and triplet scattering lengths are similar~\cite{Owens2017}.
	
	\refform{tab:ScattLengths} shows the singlet and triplet scattering lengths for different pairs of alkali metal atoms.
    These values demonstrate that \Rb{87}+\Rb{87} is a special case.
	For most other combinations of alkali-metal isotopes, the singlet and triplet scattering lengths are very different and the rate coefficients for both \rf{}-induced and \rf{}-free spin exchange will be large.
	Although \rf{}-dressed potentials may enable the manipulation of different isotopes in a mixture~\cite{Bentine2017}, this paper finds that the \rf{} dressing will generally cause high rates of inelastic collisions.
	Nonetheless, there may be some mixtures for which inelastic losses are low.
	For instance, the combinations \Lithium{6}+\Sodium{23}, \Lithium{6}+\Potassium{39} and \Lithium{6}+\Potassium{40} have similar singlet and triplet scattering lengths, which may suppress interspecies spin exchange.
	Unfortunately, the singlet and triplet scattering lengths are dissimilar in \Lithium{6}+\Lithium{6}, \Sodium{23}+\Sodium{23}, \Potassium{39}+\Potassium{39} and \Potassium{40}+\Potassium{40}, hence \rf{}-induced intraspecies spin exchange may be fast for these species.
	\Rb{87}+\Caesium{133} may also be interesting; although the interspecies scattering lengths are dissimilar, they are both large and may give rise to similar phase shifts.
	Furthermore, \Rb{87} and \Caesium{133} have different magnitudes of \gF{}, allowing species-selective manipulations.
	It is also well established that \Rb{87}+\Rb{87} has low inelastic loss rates when \rf{} dressed.
	Further calculations would be required to predict the rate coefficients for a \Rb{87}+\Caesium{133} mixture.
	
	This paper has not considered inelastic collisions between different hyperfine states of the same isotope, which can also be independently manipulated using \rf{}-dressed potentials, as was shown for \Rb{87}~\cite{Navez2016}.
	The choice of \Rb{87} was fortunate, as the similarity of singlet and triplet scattering lengths suppresses spin-exchange collisions even when such collisions are otherwise allowed~\cite{Julienne1997}.
	\rf{}-dressed potentials have found use for this specific mixture, but our work suggests that this promising technique may be more limited in scope than was previously realized.

	\acknowledgments
	
	This work has been supported by the UK Engineering and Physical Sciences Research Council
	(Grants No. ER/I012044/1, EP/N007085/1 and EP/P01058X/1 and a Doctoral Training
	Partnership with Durham University) and the EU H2020 Collaborative project QuProCS (Grant Agreement 641277).
	E.~B., A.~J.~B., K.~L.\ and D.~J.~O.\ thank the EPSRC for doctoral training funding.
	We are grateful to Dr.~C.~R.~Le Sueur and Prof.~T.~Fernholz for valuable discussions.

	\appendix{}

	\section{Numerical methods}
	\label{appendix:QSC_NumericalMethods}

We have carried out quantum scattering calculations of collisions between pairs
of atoms in \rf{}-dressed states. The Hamiltonian for the colliding pair is
\begin{align}
\frac{\hbar^2}{2\mu} \left[-R^{-1} \frac{d^2}{dR^2} R + \frac{\hat
	L^2}{R^2} \right] &+ \hat V(R) + \hat h_1 + \hat h_2 \nonumber\\
&+ \hat h_{\rm rf} + \hat h_{\rm rf,1}^{\rm int} + \hat h_{\rm rf,2}^{\rm int},
\label{eq:ham-pair}
\end{align}
where $\mu$ is the reduced mass, $\hat L^2$ is the operator for the
end-over-end angular momentum of the two atoms about one another, and $\hat
V(R)$ is the interaction operator,
\begin{equation}
{\hat V}(R) = \hat V^{\rm c}(R) + \hat V^{\rm d}(R).
\label{eq:V-hat}
\end{equation}
Here $\hat V^{\rm c}(R)=V_0(R)\hat{\cal{P}}^{(0)} + V_1(R)\hat{\cal{P}}^{(1)}$
is an isotropic potential operator that depends on the electronic potential
energy curves $V_0(R)$ and $V_1(R)$ for the singlet and triplet electronic
states and $\hat V^{\rm d}(R)$ is a relatively weak anisotropic operator that
arises from the combination of spin dipolar coupling at long range and
second-order spin-orbit coupling at short range. The singlet and triplet
projectors $\hat{ \cal{P}}^{(0)}$ and $\hat{ \cal{P}}^{(1)}$ project onto
subspaces with total electron spin quantum numbers 0 and 1 respectively. The
potential curves for the singlet and triplet states of Rb$_2$ are taken from
ref.~\onlinecite{Strauss2010}.

Expanding the scattering wavefunction in the basis set \eqref{eq:bas-u}
produces a set of coupled equations in the interatomic distance coordinate $R$.
For \Rb{87}+\Rb{85} we use a basis set with photon numbers $N$ from $-3$ to
3, $M_N=-N$, $L$ restricted to $L_{\rm max}=0$ or 2, and all possible values of
$m_{s1}$, $m_{s2}$, $m_{i1}$ $m_{i2}$ and $M_L$ that produce the required value
of the conserved quantity $M_{\rm tot}=
M_F+M_L+M_N$. The resulting number of coupled equations varies from 30 to 478.
These equations are solved using the MOLSCAT package \cite{molscat2019,mbf-github:2019}. In the
present work we use the hybrid log-derivative propagator of Alexander and
Manolopoulos \cite{Alexander1987} to propagate the coupled equations from short
range out to $R_{\rm max}=\SI{15000}{\bohr}$. At this distance,
MOLSCAT transforms the propagated solution into the asymptotic basis set and
applies scattering boundary conditions to extract the scattering matrix ${\bf
S}$. It then obtains the complex energy-dependent scattering length
$a(E,B)=\alpha(E,B)-i\beta(E,B)$ from the identity \cite{Hutson2007}
\begin{align}
a(E,B) = \frac{1}{ik} \left(\frac{1-S_{00}(E,B)}{1+S_{00}(E,B)}\right),
\label{eq:scat-length}
\end{align}
where $k^2=2\mu E/\hbar^2$ and $S_{00}(E,B)$ is the diagonal S-matrix element
in the incoming s-wave channel. For s-wave collisions (incoming $L=0$), the
rate coefficient for inelastic loss is exactly \cite{Cvitas2007}
\begin{align}
k_2(E,B) = \frac{2hg_\alpha\beta(E,B)}{\mu\left[1+k^2|a(E,B)|^2+2k\beta(E,B)\right]},
\end{align}
where $g_\alpha$ is 2 for identical bosons and 1 for distinguishable particles.
In the present work we evaluated $k_2(E,B)$ from scattering
calculations at an energy $E=0.4\ \mu{\rm K}\times k_{\rm B}$. We did not carry
out explicit energy averaging, since $k_2(E,B)$ is independent of energy in the
limit $E\rightarrow 0$. 
We verified the limit holds by performing additional calculations of $k_2$ at an energy $E=\SI{5}{\micro\K}$, and found that the inelastic rate coefficients vary by only \SI{15}{\percent} between the two sets of calculations in the field region of the experiments.

\bibliography{refs}
	
\end{document}